\documentstyle[11pt,aaspp4]{article}
\def\lya{Ly$\alpha$~}

\begin{document}

\title{Probing the Universe After Cosmological Recombination Through the
Effect of Neutral Lithium on the Microwave Background Anisotropies}

\author{Abraham Loeb}

\affil{Astronomy Department, Harvard University, 60 Garden Street,
  Cambridge, MA 02138; aloeb@cfa.harvard.edu}

\begin{abstract} 

I show that neutral lithium couples strongly to the cosmic microwave
background (CMB) through its 6708\AA~ resonant transition after it
recombines at $z\sim 500$.  At observed wavelengths of $\la 335\mu$m, the
CMB anisotropies are significantly altered since the optical depth for
resonant scattering by neutral lithium is substantial, $\tau_{\rm LiI}\sim
0.5$.  The scattering would suppress the original anisotropies, but will
generate strong new anisotropies in the CMB temperature and polarization on
sub-degree scales ($\ell \ga 100$).  Observations at different wavelengths
in this spectral regime can probe different thin slices of the early
universe. The anisotropy noise contributed by continuum radiation from
foreground far--infrared sources could be taken out by subtracting maps at
slightly different wavelengths.  Detection of the above effects can be used
to study structure at $z\la 500$ and to constrain the primordial abundance
and recombination history of lithium.

\end{abstract}


\section{Introduction}

The dynamical coupling between the baryons and the Cosmic Microwave
Background (CMB) during the epoch of cosmological recombination ($z\sim
10^3$) is commonly assumed to be mediated only through Thomson scattering
(Peebles \& Yu 1970). This coupling becomes ineffective when hydrogen
recombines, at around the same time when the universe becomes transparent,
and hence it leaves a clear signature on the temperature anisotropies of
the CMB (White, Scott, \& Silk 1994, and references therein).  Forthcoming
satellite missions, such as MAP\footnote{http://map.gsfc.nasa.gov} in 2001
or Planck\footnote{http://astro.estec.esa.nl/Planck} in 2007, offer the
sensitivity to probe deviations from this simple model at the percent level
(Scott 1999; Yu, Spergel, \& Ostriker 2001).

In this {\it Letter}, I examine the potential significance of the coupling
between the cosmic gas and the CMB due to resonant line transitions of
neutral atoms. For such transitions, the integral of the absorption
cross-section over frequency,
\begin{equation}
\int \sigma(\nu) d\nu = f_{12} \left({\pi e^2\over m_e c}\right), 
\label{eq:int_s}
\end{equation}
is typically many orders of magnitude larger than for Thomson scattering,
where $f_{12}$ is the absorption oscillator strength (Rybicki \& Lightman
1979).  For example, the \lya transition of hydrogen provides an average
cross-section which is seven orders of magnitude larger than the Thomson
value, when averaged over a frequency band as wide as the resonant
frequency itself [$(\Delta \nu/\nu) \sim 1$].  Since the abundance of atoms
in excited states is highly suppressed by large Boltzmann factors at the
recombination redshift, we consider only transitions from the ground
state. For helium, these transitions correspond to photon frequencies that
are too far on the Wien tail of the CMB spectrum and only couple to a
negligible fraction of the radiation energy density.  Hence, we focus our
discussion on hydrogen and lithium.

In \S 2, we derive the resonant--line opacity and drag force for atoms that
move relative to a blackbody radiation field.  In \S 3, we consider the
cosmological implications of these results for hydrogen and
lithium. Finally, \S 4 summarizes the main conclusions of this work.
Throughout the paper, we adopt the density parameters $\Omega_0\approx 0.3$
for matter and $\Omega_b = 0.04$ for baryons, and a Hubble constant
$H_0\approx 70~{\rm km~s^{-1}~Mpc^{-1}}$.  The effect of the cosmological
constant can be neglected at the redshifts of interest here.

\section{Drag Force and Opacity Due to a Resonant Line}

Consider a single atom moving at a non-relativistic peculiar velocity ${\bf
v}$ relative to an isotropic blackbody radiation field of temperature $T$.
Because of its motion, the atom sees a dipole anisotropy in the radiation
temperature, $T_a(\theta)=T [1+(v/c)\cos \theta]$, where $\theta$ is the
angle relative to its direction of motion (Peebles 1993).  We assume that
the atom couples to the radiation through a resonant line of frequency
$\nu_0$ with $(h\nu_0/kT)\gg 1$. Hence, the line resonates in the Wien tail
of the spectrum, for which the radiation energy density per unit frequency
(in erg~cm$^{-3}$~Hz$^{-1}$) is given by,
\begin{equation}
u_\nu(\theta)={8\pi h\nu^3\over c^3}\exp\left\{-{h\nu \over
kT_a(\theta)}\right\}\approx {8\pi h\nu^3\over c^3}\exp\left\{-{h\nu \over
kT}\right\}\left(1+ {h\nu\over kT}{v\over c}\cos \theta\right),
\end{equation}
where we have used the fact 
that $[(h\nu/kT)(v/c)\cos \theta]\ll 1$.  Absorption
of a resonant photon by the atom is followed by isotropic emission in the
atom's rest frame, and so the net drag force (momentum transfer per unit
time) is given by,
\begin{equation}
F=-\int_0^{\infty} d\nu ~~{1\over 2} \int_{-1}^{1} d \cos \theta~~
\sigma(\nu)u_\nu(\theta) \times \cos \theta~~~ .
\label{eq:int_F}
\end{equation}
Since the absorption cross-section is sharply peaked around $\nu_0$, we may
use equation (\ref{eq:int_s}) to get,
\begin{equation}
{\bf F}=-\left({8\pi^2 h^2 f_{12} \nu_0^4\over 3 c^3 kT}{e^2 \over m_e c^2}
\exp\left\{-{h\nu_0\over kT}\right\} \right) {\bf v} ~~~ .
\label{eq:force}
\end{equation}

For an ensemble of atoms of density $n_{\rm a}$, which are all moving at
the same peculiar velocity ${\bf v}$, the force per unit volume is given by
$n_{\rm a} {\bf F}$. If the resonant atoms are embedded within a hydrogen
plasma of total (free$+$bound) proton density $n_p$ and are strongly
coupled to the plasma through binary particle collisions, then the
cosmological equation of motion for the gas as a whole reads,
\begin{equation}
\left(m_p n_p + m_{\rm a} n_{\rm a}\right) \left({d{\bf v}\over dt} +H{\bf
v}\right) = n_{\rm a} {\bf F} ,
\label{eq:cosm}
\end{equation}
or equivalently,
\begin{equation}
{dv\over dt} +Hv= -{v\over t_{\rm a}},
\label{eq:acc}
\end{equation}
where $H=({\dot{a}/a})$ is the Hubble expansion rate, and $t_{\rm a}$ is
the characteristic time over which the peculiar velocity of the gas is
damped due to the drag force on the resonant atoms,
\begin{equation}
t_{\rm a}(t)\equiv {3 m_p \lambda_0^4 \over 8 \pi^2h^2 c f_{12}} {m_e
c^2\over e^2}{kT \over X_{\rm a}} \exp\left\{{h\nu_0\over kT}\right\} .
\label{eq:t_a}
\end{equation}
Here, $\lambda_0=c/\nu_0$ is the resonant wavelength, $m_p$ is the proton
mass, and $X_{\rm a} \equiv n_{\rm a}/(n_p+ A_{\rm a} n_{\rm a})$, where
$A_{\rm a}=m_{\rm a}/m_p$ is the atomic weight of the resonant atoms. In
the cosmological context, $T=2.725~{\rm K}\times (1+z)$ is the CMB
temperature (Mather et al. 1999).  Both $T$ and $X_{\rm a}$ are functions
of cosmic time, and so equation (\ref{eq:acc}) admits the solution,
$v=[{v_0/a(t)}] \exp\left\{-\int {dt /t_{\rm a}(t)}\right\}$, where
$v_0=const$ and $a(t)$ is the cosmological scale factor.  In general, the
right-hand-side of equation~(\ref{eq:acc}) should include other terms which
source the peculiar velocity field, such as gravitational or pressure
forces due to density inhomogeneities.  The significance of the drag force
can be calibrated in terms of the Hubble time, $H^{-1}(z)$, through the
product $Ht_{\rm a}$. For the redshifts of interest, $H(z)\approx
\sqrt\Omega_0 H_0 (1+z)^{3/2}$.

The assumption of an isotropic radiation background is only valid if the
moving fluid element is {\it optically-thin} to resonant scattering.  In
the opposite regime of high opacity, the background photons are isotropized
locally through multiple scatterings and exert a reduced drag force on an
atom embedded deep inside the moving fluid element.  The Sobolev optical
depth for resonant scatterings in an expanding and nearly uniform universe
is given by (Sobolev 1946; Dell'Antonio \& Rybicki 1993),
\begin{equation}
\tau_{\rm a}(z)= f_{12} {\pi e^2\over m_e c} {\lambda_0 n_{\rm a}(z) \over
H(z)}= {A_{21}(g_2/g_1) \lambda_0^3 n_{\rm a}(z)\over 8\pi H(z)}~~,
\label{eq:tau_a}
\end{equation}
where $A_{21}$ is the transition probability per unit time for spontaneous
emission between the two energy levels (in s$^{-1}$), and $(g_2/g_1)$ is
the ratio between the statistical weights of the excited and ground
states. The optical depth is dominated by the velocity gradient of the
Hubble flow, because gravitationally--induced peculiar velocities perturb
this flow only slightly at the early cosmic times of interest here.

For $\tau_{\rm a} \gg 1$, the flux of resonant photons inside the moving
fluid element is reduced by a factor of $\sim \tau_{\rm a}$ (Rybicki \&
Hummer 1978), and so the drag force in equation (\ref{eq:int_F}) is lowered
by the same factor. This reduction has dramatic consequences for the
resonant \lya transition of neutral hydrogen, as we show next.

\section{Implications For Cosmological Recombination}

\subsection{Hydrogen}

For the $1S$--$2P$ \lya transition of hydrogen, $\lambda_0=1216$\AA~,
$A_{21}=6.262\times 10^8~{\rm s}^{-1}$, $f_{12}=0.4162$, and $(g_2/g_1)=3$,
yielding an exceedingly large Sobolev optical--depth, $\tau_{\rm H}=
7.5\times 10^8 X_{\rm H} [(1+z)/10^3]^{3/2}$, where $X_{\rm H}$ denotes the
neutral fraction of hydrogen. The thermally--broadened \lya line has an
effective cross--section of $\langle\sigma\rangle\sim 3\times 10^{-17}~{\rm
cm^2}$ (Peebles 1993) and a corresponding mean-free-path of $1/n_{\rm
H}\langle \sigma\rangle \sim 10^{14}~{\rm cm}$, at $z\sim 10^3$. Only on
smaller scales can \lya viscosity damp velocity gradients effectively.  On
the much larger scales where the CMB anisotropies are measurable, the drag
time in equation (\ref{eq:t_a}) is lengthened by a factor of $\sim
\tau_{\rm H}$, which makes it very much longer than the Hubble time. The
situation is similar for deuterium. Balmer line transitions of hydrogen
have a small optical depth, but the occupation probability of their lower
level is suppressed by an exceedingly large Boltzmann factor, $\sim
\exp\{-4.35\times 10^4/(1+z)\}$.

The resulting drag force is small, even when including the enhancement in
the \lya flux beyond the blackbody value, due to the \lya photons released
during recombination ($\sim 1$~\lya photon per hydrogen atom; see
Dell'Antonio \& Rybicki 1993 for the distorted spectrum).

\subsection{Lithium}

\subsubsection{Drag Force and Opacity}

Lithium has a closed $n=1$ shell with two electrons, and one electron
outside this shell. The transition between the ground state ($2S$) and the
first excited state ($2P$) has $\lambda_0=6708$\AA~ and $A_{21}=3.69\times
10^7~{\rm s^{-1}}$, with $f_{12}=0.247$ and $(g_2/g_1)=1$ for the
$2^2S$--$2^2P^0_{1/2}$ transition, and $f_{12}=0.494$, $(g_2/g_1)=2$ for
the $2^2S$--$2^2P^0_{3/2}$ transition (Radzig \& Smirnov 1985; Yan \& Drake
1995). We therefore get from equations (\ref{eq:t_a}) and (\ref{eq:tau_a}),
\begin{equation}
H t_{\rm LiI}= 4.96\times 10^{-4} \left({1+z\over 10^3}\right)^{5/2}
\left({X_{\rm Li I} \over 3.8\times 10^{-10}}\right)^{-1}
\exp\left\{{7.89\times 10^3\over 1+z}\right\} ,
\label{eq:t_Li}
\end{equation}
and
\begin{equation}
\tau_{\rm LiI}=2.82 \left({X_{\rm Li I} \over 3.8\times 10^{-10}}\right)
\left({1+z\over 10^3}\right)^{3/2} .
\label{eq:tau_Li}
\end{equation}

Figure 1 depicts the redshift dependence of the drag time in equation
(\ref{eq:t_Li}) and compares it to the drag time due to Thomson scattering
in the recombining plasma,
\begin{equation}
H t_{\rm e}={3m_p c H\over 4\sigma_T a T^4 X_e}=
4.92\times 10^{-3} 
X_{\rm e}^{-1} \left({1+z\over 10^3}\right)^{-5/2}~~,
\end{equation}
where $X_{\rm e}=(n_{\rm e}/n_p)= n_{\rm e}/(n_{\rm e}+n_{\rm H})$ is the
ionization fraction of hydrogen. We calculated $X_{\rm e}(z)$ from the
updated version of the standard recombination model (Seager, Sasselov, \&
Scott 2000).  Despite the low abundance of lithium, its resonant drag force
is not much lower than the Thomson drag force at $z\sim 500$, when roughly
half of the lithium atoms are expected to have recombined (Palla, Galli, \&
Silk 1995; Stancil, Lepp, \& Dalgarno 1996, 1998).

The net effect of the lithium drag is highly sensitive to the recombination
history of lithium and hydrogen.  Previous discussions (e.g., Stancil et
al.  1996, 1998) did not include the delaying effect of \lya photons on the
recombination history of hydrogen (Peebles 1993), and therefore
underestimated the free electron fraction and correspondingly the neutral
lithium abundance at $z\la 10^3$.  Another process with an opposite sign
that was omitted, involves the excitation of neutral lithium from its
ground level by the CMB radiation field.  For this effect, it is important
to take account of the Lyman and Balmer series distortions of the CMB
spectrum (Dell'Antonio \& Rybicki 1993), which redshift into resonance at
lithium recombination. In fact, the lithium chemistry in the early universe
might provide a way for inferring the existence of these distortions.
Finally, the recombination history of $^7$Be should also be followed since
a substantial fraction of the $^7$Li forms through electron-capture in the
$^7$Be produced by Big-Bang nucleosynthesis (T. Walker, private
communication). Since the half-life of neutral $^7$Be is only 53 days, this
conversion can be regarded as instanteneous on the cosmological timescale,
as soon as $^7$Be recombines. Careful calculations of the neutral lithium
fraction as a function of redshift are required in order to quantify the
imprint of the lithium drag on the CMB anisotropies (Dalgarno, Loeb, \&
Stancil 2001).

The lithium chemistry could also be affected by a large drift velocity
between the Li I fluid and the hydrogen plasma.  To examine this effect, we
consider the equation of motion for the Li I fluid (Burgers 1969),
\begin{equation}
m_{\rm LiI}
\left( {d{\bf v}_{\rm LiI}\over dt} + H{\bf v}_{\rm LiI}\right)=
-m_{\rm LiI} {{\bf v}_{\rm LiI}\over X_{\rm LiI} t_{\rm LiI}}
+ m_{\rm p} W_{\rm LiI}
\left({\bf v} - {\bf v}_{\rm LiI}\right)~~.
\label{eq:Li_eom}
\end{equation}
Here $W_{\rm LiI}=\Sigma_k \mu_{\rm k:Li} W_{\rm k:Li}$ is the collision
rate of a Li atom with the gas, where $W_{\rm k:LiI}=n_{\rm k}\langle
\sigma_{\rm k:LiI} v\rangle$, and $\mu_{\rm k:Li}=[m_{\rm k}m_{\rm
Li}/m_{\rm p}(m_{\rm k}+m_{\rm Li})]$. The index ${\rm k} $ runs over all
other particle species in the gas which are assumed to move at a common
bulk velocity ${\bf v}$ because they are not subject to the resonant drag
force.  The geometric cross-section of Li I atoms, $\sigma_{\rm LiI} \sim
10^{-15}~{\rm cm^2}$, couples them to the rest of the gas on a time scale
$\sim 10^7~{\rm s}$, much shorter than the Hubble time, $H^{-1}\sim 3
\times 10^{13}~{\rm s}$ at $z\sim 10^3$. Hence, we may ignore the terms on
the left-hand-side in equation (\ref{eq:Li_eom}), as well as the perturbed
gravitational force which is of the same order. The remaining terms give
the solution,
\begin{equation}
{\left(v -v_{\rm LiI}\right)\over v}={1\over 1+ Y} 
\end{equation}
where $Y= W_{\rm LiI} t_{\rm LiI} X_{\rm LiI}/A_{\rm Li}$.  Substituting
$A_{\rm Li}=7$, we find that $W_{\rm LiI} \approx
10^{-7}[(1+z)/10^3]^{3.5}~{\rm s^{-1}}$, and that $Y\gg 1$ at $z\la 400$,
after lithium recombination.  At higher redshifts, the assumption that
$v_{\rm LiI}=v$ seems not to apply and one may wonder whether
the drag time in equation (\ref{eq:acc}) should be multiplied by a factor
of $(v/v_{\rm LiI})=(1+Y)/Y > 1$, since the drag force acts on
${\bf v}_{\rm Li I}$ and not on ${\bf v}$. In this context it is crucial to
realize that the Li I nuclei do not maintain their identity as neutral
atoms; in fact, they continuously form and get ionized on a timescale much
shorter than a Hubble time. The cross-section for Coulomb collisions of
Li$^+$ is far greater than that of Li I, $\sigma_{\rm Li^+} =2\sqrt\pi
e^4\ln \Lambda/(kT)^2 = 2.7 \times 10^{-11} [(1+z)/10^3]^{-2}~{\rm cm^2}$,
where $\ln \Lambda\approx 20$ is the Coulomb logarithm.  The fraction of
time spent by a Li nucleus in an ionized state is the ionization fraction,
$f_{+}=X_{\rm Li^+}/(X_{\rm LiI}+X_{\rm Li^+})$, and so $\sigma_{\rm LiI}$
should be replaced by $[f_+\sigma_{\rm Li^+}+ (1-f_+)\sigma_{\rm LiI}]$ and
the drag force term should be multiplied by $(1-f_+)$ in equation
(\ref{eq:Li_eom}). We consequently find that the substantial ionization
fraction of the lithium fluid (Stancil et al. 1996, 1998) diminishes its
drift velocity relative to the gas at all relevant redshifts, $z\la
10^3$. Hence, non-thermal relative velocities may be ignored in the Li I
chemistry.

At $z\la 10^3$, the mean-free-path for resonant scattering by lithium
exceeds the relevant spatial scales for the CMB anisotropies
($10^{23}$--$10^{24}~{\rm cm}$), even if the lithium abundance is taken to
be somewhat higher than the value, $X_{\rm SBBN}=3.8\times 10^{-10}$,
predicted by the latest deuterium measurements and Standard Big-Bang
Nucleosynthesis (Burles, Nollett, \& Turner 2001; see also Walker et
al. 1991; Smith, Kawano, \& Malaney 1993). Note that values as high as
$X_{\rm Li I}\sim 10^{-8}$ were suggested by models of inhomogeneous
Big-Bang nucleosynthesis (Applegate \& Hogan 1985; Sale \& Mathews 1986;
Mathews et al. 1990), and would result in $Ht_{\rm LiI}\la 10$ for velocity
gradients on sufficiently small scales with $\tau_{\rm LiI}\la 1$.

\subsubsection{Effects on CMB Anisotropies}

If a substantial fraction of lithium recombines by $z\sim 500$ (as
predicted by Palla et al. 1995 or Stancil et al. 1996, 1998), then the flux
of the original anisotropies will be suppressed by the absorption factor,
$\exp (-\tau_{\rm LiI})$, at observed wavelengths below $(6708$\AA~$\times
500)=335\mu$m, for which $\tau_{\rm LiI}\approx 0.6 (X_{\rm Li I}/X_{\rm
SBBN})$.  The resonant scattering effect is different from that caused by
Thomson scattering at the epoch of reionization (Hu \& White 1997), in that
it has a larger optical depth and it occurs within a much thinner shell of
gas (for a sufficiently narrow band of observed wavelengths).
Consequently, it should induce new first-order anisotropies [multiplied by
$(1-\exp (-\tau_{\rm LiI})$] due to the coherence of the velocity field in
the thin scattering shell.  The dominant contribution to the new
anisotropies would come from the Doppler effect.  In contrast to
reionization, there should be no severe cancellations of the Doppler
effects from line-of-sight velocity fluctuations at large wavenumbers,
because the last scattering surface (or ``visibility function'') for the
resonance is extremely thin.  The increase in amplitude of sub-horizon
fluctuations between $z=10^3$ and $z=500$ would enhance the new
anisotropies relative to the original ones and introduce a distortion of
order unity in the original anisotropy spectrum on these scales.  The
anisotropy spectrum would be modified at multipole moments $\ell \ga 100$,
around and below the angular scale of the first acoustic peak (which is a
factor of $\sim \sqrt{6}=2.4$ smaller than the horizon scale at $z\sim
500$), where the original anisotropy amplitude reaches its maximum ($\ell
\approx 220$).  Resonant scattering would also result in enhanced
polarization anisotropies on the same angular scales (Chandrasekhar 1960).
Since the scattering is done by lithium, the anisotropies will reflect the
peculiar velocity of the lithium fluid which, as shown above, is not
expected to deviate significantly from the velocity of the gas as a whole.

Since the temperature fluctuations are in the Wien tail of the CMB
spectrum, they translate to brightness fluctuations (in ${\rm
erg~s^{-1}cm^{-2}sr^{-1}Hz^{-1}}$) of a much larger contrast,
\begin{equation}
{\Delta B_\nu\over B_\nu}= 
\left({d\ln B_\nu \over d\ln T}\right){\Delta T\over T}=
\left({h\nu\over kT}\right){\Delta T\over T}= 
15.78 \times \left({500\over 1+z}\right){\Delta T\over T}~~,
\end{equation}
where $z$ is the redshift being probed and we have substituted
$B_\nu(T)\propto \exp(-h\nu/kT)$. Hence the first acoustic peak would
correspond to brightness fluctuations of up to $\sim 5\times 10^{-4}$.

At a given observed wavelength, the redshift thickness of the scattering
shell is only determined by two contributions: (i) the possibility that a
photon at this wavelength was either absorbed by the $2^2S$--$2^2P^0_{3/2}$
transition with $\lambda_0=6707.76$\AA~or at a slightly lower redshift by
the $2^2S$--$2^2P^0_{1/2}$ transition with $\lambda_0=6707.91$\AA~(Radzig
\& Smirnov 1985), yielding $[\Delta z/(1+z)] =2.2\times 10^{-5}$; and (ii)
the thermal width of the line\footnote{Since $\tau_{\rm LiI}\la 1$, the
photon does not need to redshift away from the line center by more than the
line width in order to escape to infinity.}, $[\Delta z/(1+z)] = (\Delta
\lambda/\lambda)_{\rm th} \approx 10^{-5}$. In practice, the minimum
thickness would be limited by the band--width of the detector.  The maximum
band--width and minimum angular resolution should be tuned to match the
expected coherence length of the peculiar velocity field at $z\sim 500$,
which corresponds to $(\Delta \lambda/\lambda)\sim 0.2$ and $\Delta
\theta\sim 6^{\prime}$ or equivalently $\ell\sim 10^3$.

Observations at different far--infrared (FIR) wavelengths would probe
different thin slices of structure in the early universe.  The anisotropy
pattern would vary gradually as a function of wavelength and ``3D
tomography'' is in principle possible, by which one may probe correlations
in the structure along the line-of-sight. The feasibility of this
measurement might be severely compromised by the anisotropies of the FIR
foreground.  Unfortunately, $350\mu$m is just the wavelength where the CMB
intensity blends into the extragalactic FIR foreground (see Fig. 1 in Scott
et al. 2001). However, the contribution of discrete sources (see Fig. 6 in
Knox et al. 2001) could be separated out through observations at shorter
wavelengths with a higher angular resolution. Furthermore, continuum
foreground emission would result in similar anisotropies for all
wavelengths within a band width $(\Delta \lambda/\lambda)\sim 1$, while the
CMB anistropies would vary considerably across such a band since its width
corresponds to $\Delta z/(1+z)\sim 1$. Subtraction of anisotropy maps at
slightly different wavelengths within such a band can therefore be used to
isolate out the CMB component (as long as these wavelengths do not overlap
with strong emission lines from the foreground sources).  For $\ell\la
100$, the anisotropies should have the same power-spectrum as measured by
MAP or Planck at much longer wavelengths.  The predictable correlations
between the temperature and polarization patterns of the CMB and their
general statistical properties could be used to further enhance the
signal-to-noise ratio (Zaldarriaga \& Loeb 2001).  Obviously, these
difficult observations need to be made through holes in the Galactic dust
emission at high Galactic latitudes.

\section{Conclusions} 

The 6708\AA~ resonant transition of neutral lithium provides a previously
unexplored coupling between the baryons and the CMB after cosmological
recombination.  The lithium opacity is substantial at observed wavelengths
$\la 335 [(1+z_{\rm LiI})/500]\mu$m, where $z_{\rm LiI}$ is the redshift at
which $\sim 50\%$ of the lithium recombines. At these wavelengths, the CMB
anisotropies on sub-degree angular scales would be significantly altered,
due to the finite optical depth for resonant scattering, $\tau_{\rm
LiI}\approx 0.5[(1+z_{\rm LiI})/500]^{3/2}$.  This scattering would
generate new temperature and polarization anisotropies at mutipole moments
$\ell \ga 100$. The detection of these anisotropies depends on the
prospects for cleaning the expected anisotropies of the FIR foreground
(Knox et al. 2001), and may be feasible through the subtraction of
anisotropy maps for pairs of slightly different wavelengths.

The relevant wavelength range overlaps with the highest frequency channel
of the Planck mission ($352\mu$m) and with the proposed baloon--borne
Explorer of Diffuse Galactic
Emissions\footnote{http://topweb.gsfc.nasa.gov} (EDGE) which will survey
1\% of the sky in 10 wavelength bands between $230$--$2000\mu$m with a
resolution ranging from $6^\prime$ to $14^\prime$ (see Table 1 in Knox et
al. 2001). However, a new instrument design with multiple narrow bands
($\Delta \lambda/\lambda\la 0.1$) at various wavelengths in the range
$\lambda=250$--$350\mu$m and with angular resolution $\sim 6^\prime$, is
necessary in order to optimize the detection of the lithium signature on
the CMB anisotropies.

The effects discussed in this {\it Letter} depend critically on the
recombination history of lithium, and emphasize the need to explore the
related lithium chemistry in more detail in the future.

\acknowledgements

The author thanks Alex Dalgarno, Daniel Eisenstein, George Rybicki, 
and Matias Zaldarriaga for useful discussions.  This work was supported in
part by NASA grants NAG 5-7039, 5-7768, and by NSF grants AST-9900877,
AST-0071019.

\newpage

\begin{figure}
\begin{center}
\plotone{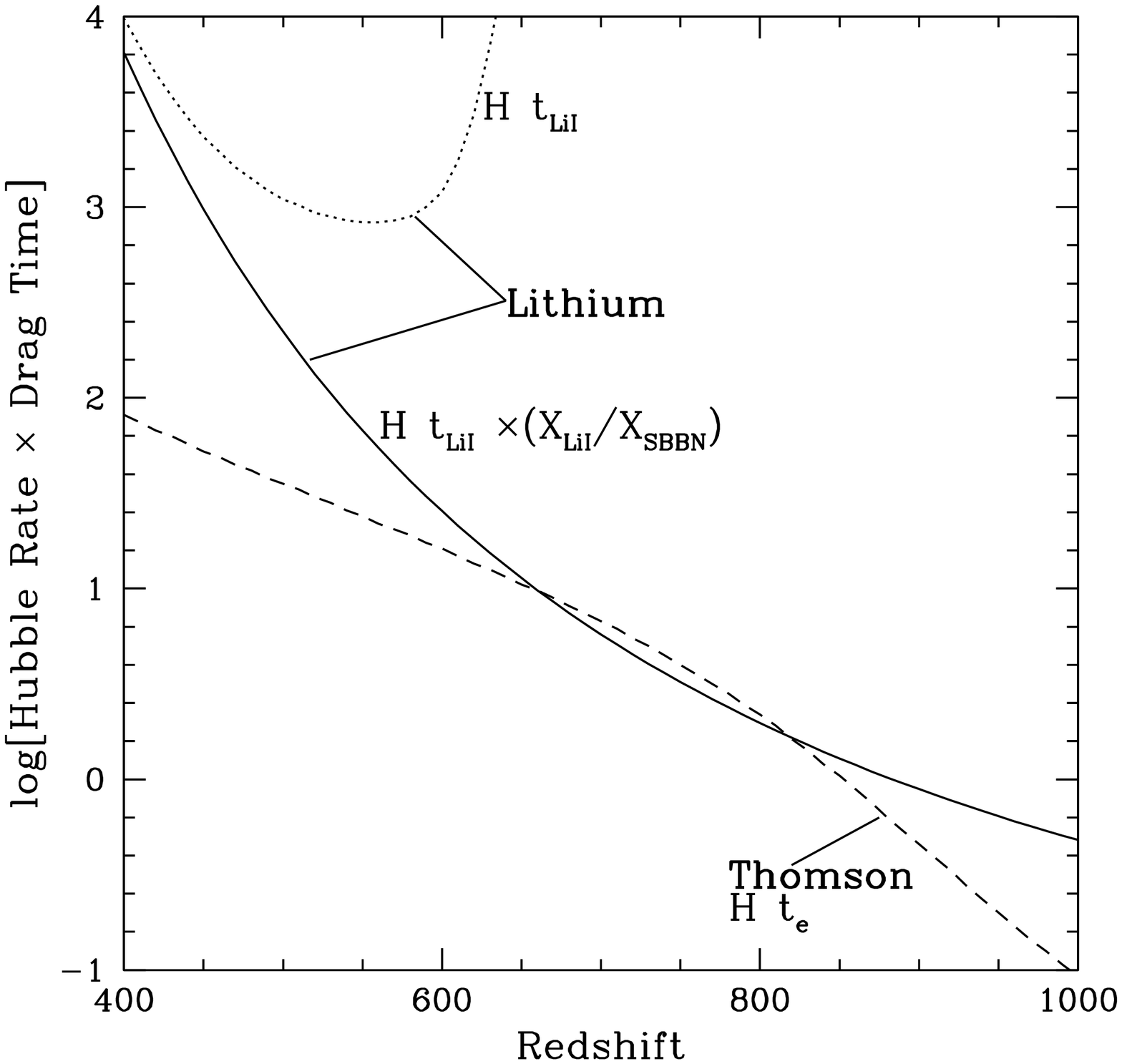} 

\figcaption[fig1]{Redshift evolution of the dynamical drag time due to the
Li 6708\AA~ resonance with the microwave background. The drag time shown by
the solid line should be divided by the neutral fraction of the primordial
lithium abundance ($X_{\rm SBBN}=3.8\times 10^{-10}$, see Burles et
al. 2001), as a function of redshift. The dotted line shows the result for
one of the recombination histories of lithium in the literature (Palla et
al. 1995).  For comparison, the Thomson drag time is plotted as the dashed
line, taking account of the full recombination history of hydrogen. In all
cases, the drag times are normalized by the Hubble time, $H^{-1}(z)$.}

\label{Fig1}
\end{center}
\end{figure}

\end{document}